\documentclass[fleqn,10pt]{wlscirepb}
\title{Crossover from a heavy fermion to intermediate valence state in noncentrosymmetric Yb$_2$Ni$_{12}$(P,As)$_7$}
\author[1]{W. B. Jiang}
\author[1]{L. Yang}
\author[1]{C. Y. Guo}
\author[2]{Z. Hu}
\author[3]{J. M. Lee}
\author[1]{M. Smidman}
\author[1]{Y. F. Wang}
\author[1]{T. Shang}
\author[1]{Z. W. Cheng}
\author[1]{F. Gao}
\author[3]{H. Ishii}
\author[3]{K. D. Tsuei}
\author[3]{Y. F. Liao}
\author[1,4]{X. Lu}
\author[2]{L. H. Tjeng}
\author[3,*]{J. M. Chen}
\author[1,4,$\dag$]{H. Q. Yuan}
\affil[1]{Center for Correlated Matter and Department of Physics, Zhejiang University, Hangzhou, 310058, China}
\affil[2]{Max Planck Institute for Chemical Physics of Solids, D-01187 Dresden, Germany}
\affil[3]{National Synchrotron Radiation Research Center, Hsinchu 30076, Taiwan}
\affil[4]{Collaborative Innovation Center of Advanced Microstructures, Nanjing 210093, China}
\affil[*]{jmchen@nsrrc.org.tw}
\affil[$\dag$]{hqyuan@zju.edu.cn}

\begin{abstract}
We report measurements of the physical properties and electronic structure of the hexagonal compounds Yb$_2$Ni$_{12}Pn_7$ ($Pn$=P, As) by measuring the electrical resistivity, magnetization, specific heat and partial fluorescence yield x-ray absorption spectroscopy (PFY-XAS). These demonstrate a crossover upon reducing the unit cell volume, from an intermediate valence state in Yb$_2$Ni$_{12}$As$_7$ to a heavy-fermion paramagnetic state in Yb$_2$Ni$_{12}$P$_7$, where the Yb is nearly trivalent. Application of pressure to Yb$_2$Ni$_{12}$P$_7$ suppresses $T_{FL}$, the temperature below which Fermi liquid behavior is recovered, suggesting the presence of a quantum critical point (QCP) under pressure. However, while there is little change in the Yb valence of Yb$_2$Ni$_{12}$P$_7$ up to 30 GPa, there is a strong increase for Yb$_2$Ni$_{12}$As$_7$ under pressure, before a near constant value is reached. These results indicate that any magnetic QCP in this system is well separated from strong valence fluctuations. The pressure dependence of the valence and lattice parameters of Yb$_2$Ni$_{12}$As$_7$ are compared and at 1~GPa, there is an anomaly in the unit cell volume as well as a change in the slope of the Yb valence, indicating a correlation between structural and electronic changes.
\end{abstract}
\begin{document}

\maketitle

\section*{Introduction}

Strongly correlated, rare-earth based intermetallic compounds have attracted considerable interest due to the wide range of novel behaviors which are often observed. The ground state of many of these systems can be tuned by adjusting non-thermal parameters, which alter the relative strengths of the Kondo and Ruderman-Kittel-Kasuya-Yosida (RKKY) interactions.\cite{Doniach} For Ce-based compounds, pressure can drive the system from magnetic ordering where the RKKY interaction dominates, to an intermediate valence (IV) state where there is strong hybridization between the $4f$ and conduction electrons. In many systems, the ordering temperature can be tuned to zero at a quantum critical point (QCP) and in this region, non-Fermi liquid (NFL) behavior and unconventional superconductivity (SC) are often observed.\cite{NatureMagSC} Although superconductivity in the vicinity of the QCP is believed to be mediated by spin fluctuations, there has been particular attention to the role played by fluctuations of the rare-earth valence in determining the ground state properties. For example, the temperature-pressure phase diagram of CeCu$_2$(Si$_{1-x}$Ge$_x$)$_2$ exhibits two superconducting domes,\cite{HQY2003} where the low-pressure phase is in close proximity to the suppression of magnetic order, while it has been proposed that the high-pressure SC phase is mediated by critical valence fluctuations.\cite{Miyake2004,HQY2006} A pair of superconducting domes, one associated with spin and the other with valence fluctuations has also been suggested to explain the superconductivity of PuCoIn$_5$ and PuCoGa$_5$, where the lattice of PuCoIn$_5$ is significantly larger than PuCoGa$_5$ and the $5f$ electrons display more localized behavior.\cite{PuCoGa5}

Generally, Yb with an electronic configuration $4f^{13}$ can be considered to be a hole counterpart of Ce with $4f^1$ . Pressure has the opposite effect to Ce compounds, since it reduces the strength of the Kondo interaction, leading to magnetic ordering with the Yb in the Yb$^{3+}$ state. While superconductivity was found in many Ce-based systems, to date the only Yb-based heavy fermion superconductor reported is $\beta$-YbAlB$_4$, which undergoes a SC transition at a low temperature of $T_c$~=~0.08~K. \cite{Nakatsuji2008} This compound is itself close to a QCP and displays NFL behavior, without the need to tune parameters such as pressure or magnetic field.\cite{Nakatsuji2008,Matsumoto} This behavior is accompanied by strong valence fluctuations, with an Yb valence of +2.75 at 20~K,\cite{Okawa2010} despite the localized nature of the magnetic moment.\cite{Macaluso2007} There is therefore particular interest in understanding the relationship between valence fluctuations and quantum criticality in Yb-based compounds and the effect of this interplay on the physical properties.

\begin{figure}[tb]
     \centering
     \includegraphics[width=0.4\linewidth]{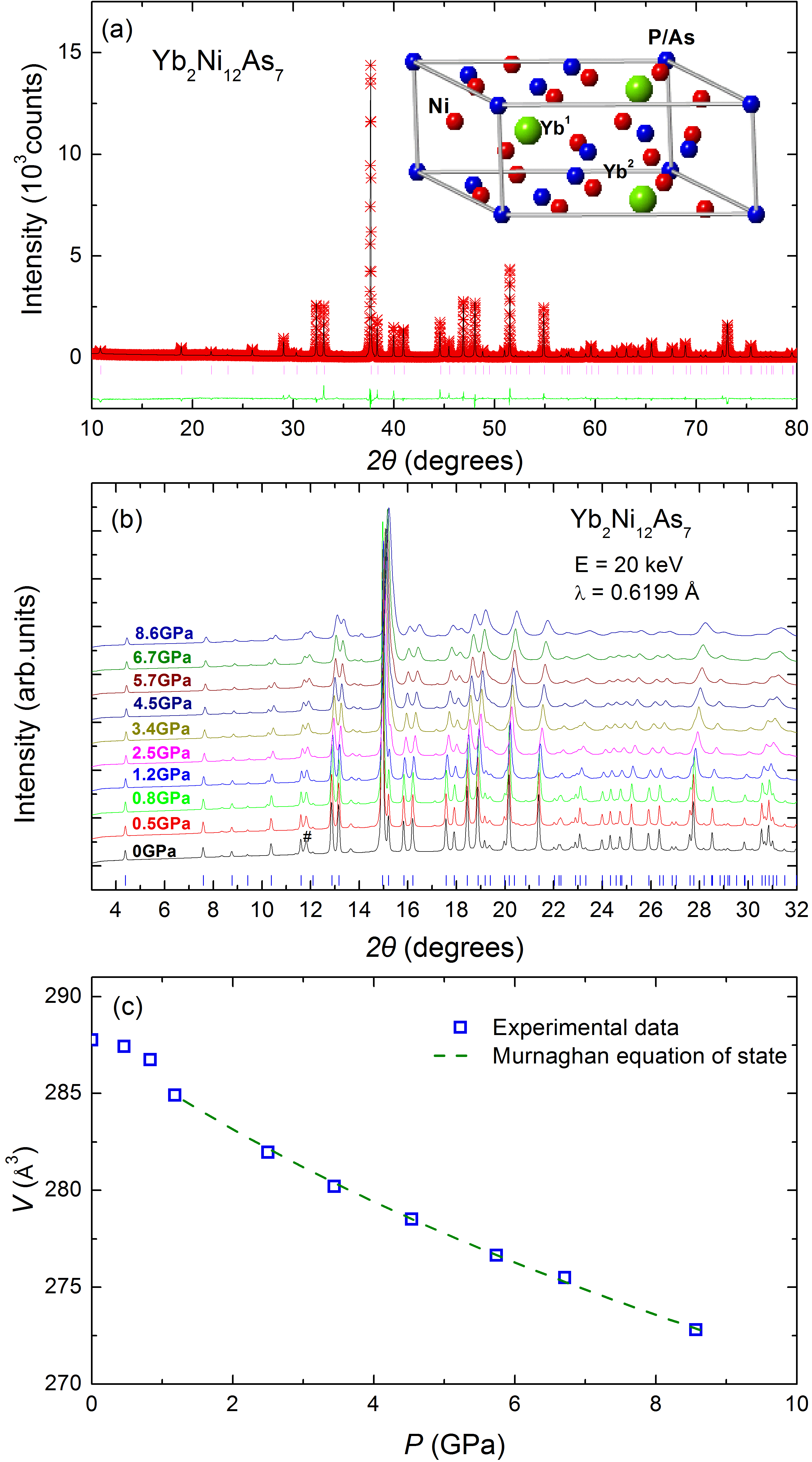}
     \caption{(Color online) (a) Powder x-ray diffraction pattern of polycrystalline Yb$_2$Ni$_{12}$As$_7$. The red crosses and black solid line denote the experimental data and calculated profiles, respectively, while the vertical bars indicate the theoretical Bragg peak positions. The crystal structure of the compounds is shown in the inset.(b) Powder x-ray diffraction patterns of Yb$_2$Ni$_{12}$As$_7$ measured under applied hydrostatic pressure. (c) Pressure dependence of the unit cell volume, obtained from refinements of the XRD data. The dashed line shows a fit to the Murnaghan equation of state, as described in the text.}
     \label{fig1}
\end{figure}

$R_2T_{12}$$Pn_7$ ($R$~=~rare earth elements, $T$~=~transition metal, $Pn$~=As or P) is a large family of compounds which crystallize in the non-centrosymmetric Zr$_2$Fe$_{12}$P$_7$-type structure with the space group P$\bar{6}$. In the unit cell, the two rare-earth atoms occupy inequivalent positions in voids of the derived Cr$_{12}$P$_7$-type structure. A variety of physical properties have been observed in these compounds, such as magnetism, multipolar order, quantum criticality and valence fluctuations.  For example, Yb$_2$Co$_{12}$P$_7$ exhibits two magnetic transitions, one at 136~K due to the ferromagnetic ordering of the cobalt-sublattice and an additional field-induced ferromagnetic transition at 5~K, which is likely associated with the ordering of the Yb ions.\cite{Hamlin} For Yb$_2$Fe$_{12}$P$_7$, an unusual temperature-field phase diagram has been proposed, where a magnetic QCP separates a low-field magnetic ordered NFL region and a high-field NFL state.\cite{Baumbach2010}

\begin{figure}[t]
     \begin{center}
     \includegraphics[width=0.4\linewidth]{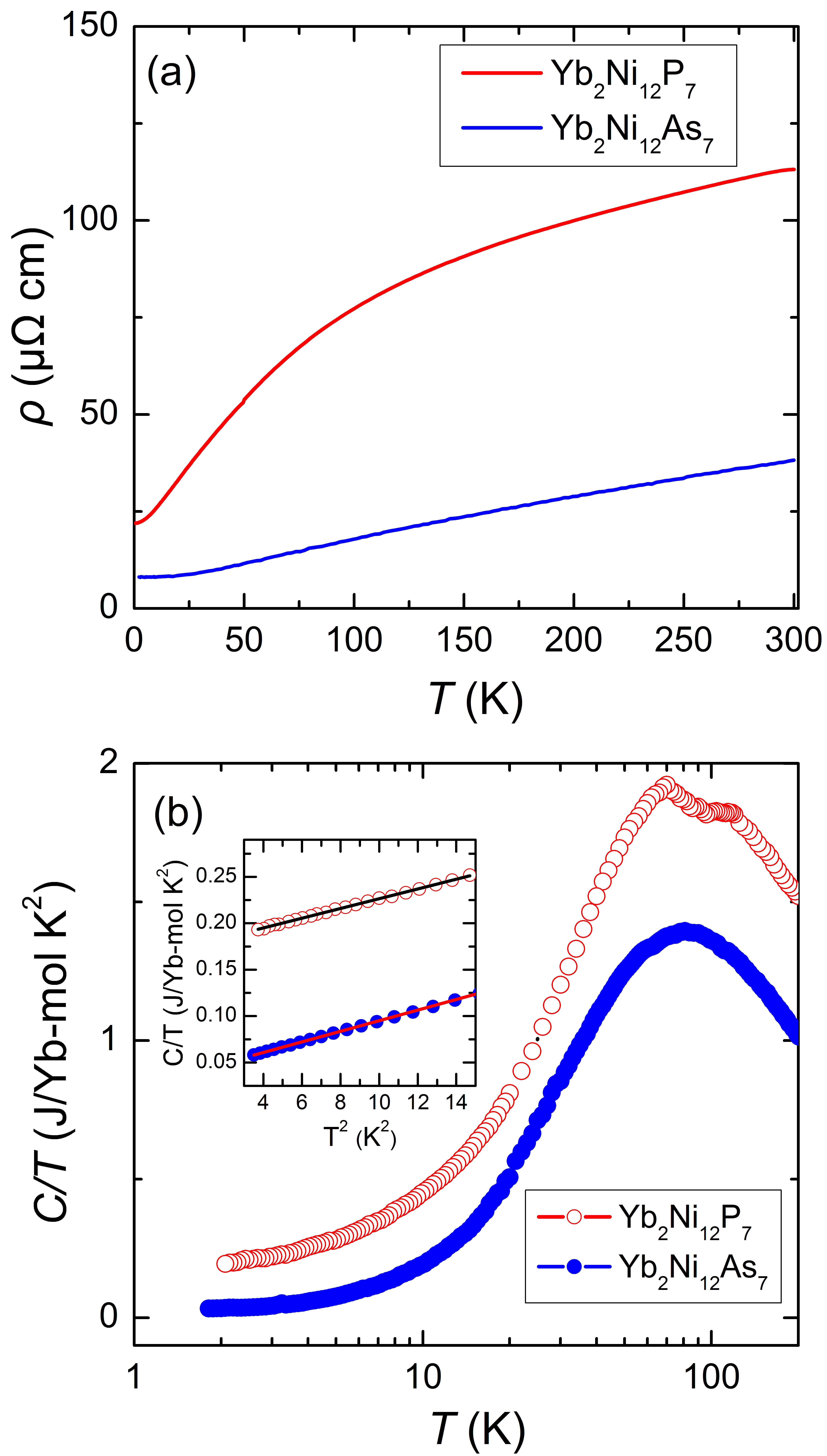}
     \end{center}
     \vspace{-12pt} \caption{(Color online) Temperature dependence of (a) the electrical resistivity $\rho(T)$ and (b) the specific heat $C/T$ of Yb$_2$Ni$_{12}$P$_7$ and Yb$_2$Ni$_{12}$As$_7$. The inset of (b) shows $C/T$ against $T^2$ at low temperatures, which are fitted using $C/T=\gamma+\beta T^2$, where $\gamma T$ and $\beta T^3$ are the electronic and lattice contributions, respectively.}
     \label{fig2}
\end{figure}

Another example is Yb$_2$Ni$_{12}$P$_7$ (YNP), which was initially believed to be an intermediate valence (IV) compound \cite{Cho}, although it has also been described as a heavy fermion compound with an enhanced electronic specific heat coefficient of $\gamma=390$ mJ/Yb-mol~K$^2$ .\cite{Nakano2012} The Yb valence is calculated to be +2.79 from the interconfiguration fluctuation (ICF) model, based on fitting magnetic susceptibility measurements. \cite{Cho} Recently, it was also reported that there is a crossover from NFL behavior at high temperatures to a Fermi liquid (FL) ground state. \cite{Jang} When the pressure is increased, the temperature below which there is an onset of FL behavior ($T_{FL}$) decreases.\cite{Tomohito} Stronger valence fluctuations were proposed in the isostructural Yb$_2$Ni$_{12}$As$_7$ (YNA), based on the larger unit cell,  lower value of $\gamma=100$ mJ/Yb-mol~K$^2$ and a higher temperature of the broad maximum in the magnetic susceptibility. \cite{Cho} Nevertheless, as of yet the presence of mixed valence behavior has not been directly measured in either of these compounds.

In this article, we report detailed measurements of the resistivity, magnetic susceptibility and specific heat of single crystals of Yb$_2$Ni$_{12}$P$_7$ and Yb$_2$Ni$_{12}$As$_7$. We have also measured the pressure and temperature dependence of the Yb valence of polycrystalline samples of these two compounds, using the bulk-sensitive partial fluorescence yield x-ray absorption spectroscopy (PFY-XAS). The pressure dependencies of the crystal structure of Yb$_2$Ni$_{12}$As$_7$ and the resistivity of Yb$_2$Ni$_{12}$P$_7$ have also been obtained, allowing the changes of the Yb valence to be related to the corresponding physical properties.

\section*{Results}

\subsection*{Crystal structure}

Figure~\ref{fig1}(a) shows powder XRD patterns at ambient conditions for polycrystalline YNA. The diffraction peaks are consistent with the calculated pattern of Yb$_2$Ni$_{12}Pn_7$, with only a few small unindexed peaks from a small quantity of Yb$_2$O$_3$ impurity. The patterns are consistent with a Zr$_2$Fe$_{12}$P$_7$-type hexagonal structure with space group P$\bar{6}$ (No.174). The powder XRD patterns for polycrystalline YNP (not displayed) also show that the samples are single phase. The crystal structure of the single crystals was also checked by powdering and measuring XRD and similar patterns were obtained. The chemical composition of single crystals were measured to be $2:12:7$ by energy-dispersive x-ray spectroscopy (EDX). To investigate the possibility of a structural phase transition under pressure, powder XRD measurements were also performed on YNA as shown in Fig.~\ref{fig1}(b), which shows no evidence of a structural phase transition up to at least 8.6~GPa. The broadening of the diffraction peaks at high pressure may be due to a reduction of the particle size, as well as an inhomogenous pressure distribution in the sample. Fig.~\ref{fig1}(c) shows the fitted unit cell volume as a function of applied physical pressure, which is obtained by refining the high pressure XRD data shown in Fig.~\ref{fig1}(b). A pronounced decrease in the unit cell volume is observed at around 1~GPa, whereas at higher pressures, the volume decreases smoothly. In order to estimate the pressure dependence of the unit cell volume at higher pressures, the data above 1.2~GPa were fitted with the expression for the Murnaghan equation of state,\cite{HighPBook}$P(V)~=~(B_0/B'_0)[(V_0/V)^{B'_0}-1]$, where the fitted parameters $B_0~=~111.3$~GPa, $B'_0~=~13.2~$GPa and $V_0~=287.76~$\AA$^{3}$ are the bulk modulus, first derivative of the bulk modulus and unit cell volume at zero pressure respectively.

\subsection*{Physical properties in ambient conditions}

The temperature dependence of the electrical resistivity [$\rho(T)$] of single crystals of YNP and YNA is shown in Fig.~\ref{fig2}(a). For YNP, $\rho(T)$ decreases with decreasing temperature and shows a broad hump around 100~K, which is likely due to hybridization between $4f$ and $5d$ electrons, as is typically observed in Kondo compounds. At low temperatures, $\rho(T)$ shows Fermi liquid behavior with $\rho(T)=\rho_0+AT^2$, where a residual resistivity $\rho_0$~=~21.96~$\mu$$\Omega$~cm and $A$~=~0.047$\mu$$\Omega$ cm/K$^2$ are obtained. No magnetic or superconducting transitions are observed down to 0.3~K.  In YNA, $\rho(T)$ shows a simple metallic behavior down to $2$~K, with no evidence for any phase transitions and the smaller resistivity indicates reduced Kondo scattering compared to YNP. The specific heat ($C/T$) from $2-200$~K of single crystals of YNP and YNA is shown in Fig.~\ref{fig2}(b). The low temperature specific heat of both compounds can be fitted with $C =\gamma T+\beta T^3$, as shown in the inset, where $\gamma T$ and $\beta T^3$ are the electronic and lattice contributions respectively. The fitted values of the electronic specific heat coefficient $\gamma$ are $173.9$~mJ/Yb-mol K$^2$ and $37.4$~mJ/Yb-mol K$^2$ for YNP and YNA respectively, while the respective Debye temperatures ($\theta_D$) are 157~K and 153~K using $\theta_D=\sqrt[3]{12\pi^4nR/5\beta}$, where $n$ is the number of atoms per formula unit and $R$ is the molar gas constant. The Kadowaki-Woods ratio ($R_{KW}$) of YNP, defined as $A/\gamma^2$, is 1.6 $\times$ 10$^{-6}$ $\mu$$\Omega$cm$\cdot$(K$\cdot$Yb-mol/mJ)$^2$, which is compatible with the previous results \cite{Jang}and is close to that calculated by using 10$^{-5}/(N(N-1)/2)$ $\mu$$\Omega$cm$\cdot$(K$\cdot$mol/mJ)$^2$ with an orbital degeneracy of $N=4$ for the $f$ electrons. \cite{Tsuji2005} Observations of a Fermi-liquid ground state, a large value of $\gamma$ and a value of $R_{KW}$ close to the expected value for heavy fermion systems, indicate that YNP is a heavy fermion compound with strong hybridization between $4f$ and conduction electrons.

\begin{figure}[t]
     \begin{center}
     \includegraphics[width=0.4\linewidth]{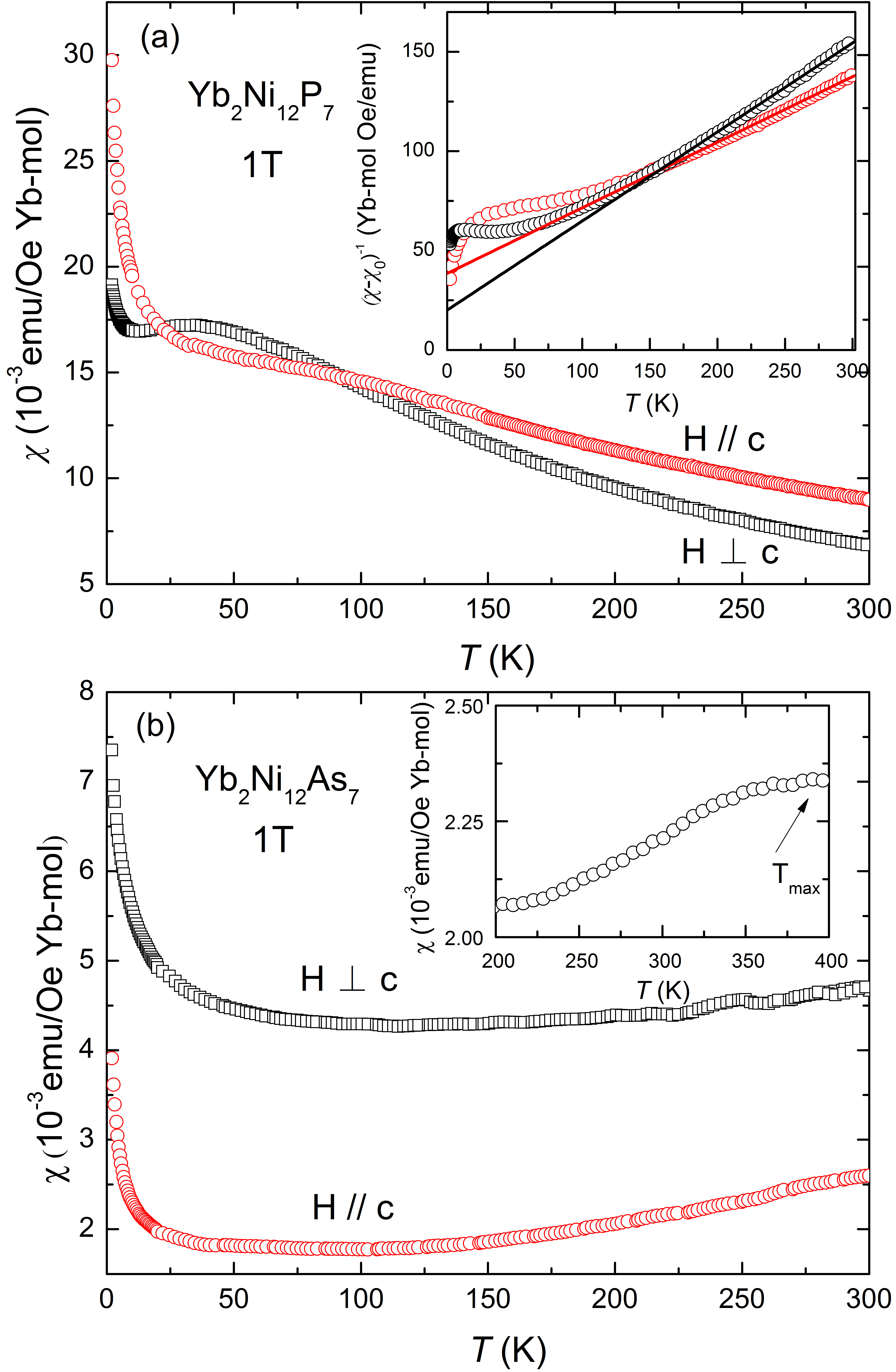}
     \end{center}
     \vspace{-12pt} \caption{(Color online) Temperature dependence of the magnetic susceptibility $\chi(T)$ for single crystals of (a) Yb$_2$Ni$_{12}$P$_7$ and (b) Yb$_2$Ni$_{12}$As$_7$ in an applied field of $1~$T which was applied both parallel and perpendicular to the $c$~axis. The inset of (a) shows the inverse susceptibility of Yb$_2$Ni$_{12}$P$_7$, which has been fitted using the modified Curie-Weiss formula, while the inset of (b) shows the magnetic susceptibility of polycrystalline Yb$_2$Ni$_{12}$As$_7$ from 200-400~K.}
     \label{fig3}
\end{figure}

Figure~\ref{fig3} shows the temperature dependence of the magnetic susceptibility for single crystals of both compounds, with an applied magnetic field of $1$~T for both $H\parallel c$ and $H\bot c$. For YNP, there is a broad maximum at around 50~K for $\chi^{H\bot c}(T)$, which may arise from the interplay of the Kondo interaction and the crystalline electric field splitting of the Yb ground state multiplet. At high temperatures above 150~K, both $\chi^{H\bot c}(T)$ and $\chi^{H\parallel c}(T)$ can be fitted using the modified Curie-Weiss law $\chi=\chi_0+C/(T-\theta_p)$, where $\chi_0$ is the temperature-independent contribution to magnetic susceptibility, $C$ is the Curie constant and $\theta_p$ is the Curie-Weiss temperature. The fitting for the inverse susceptibility is shown by the solid lines in the inset of Fig.~\ref{fig3}(a), which gives respective values of the effective magnetic moment $\mu_{eff}$ and $\theta_p$ of 4.89$\mu_B$ and -117~K for $H \parallel c$, 4.22$\mu_B$ and -44.78~K for $H \perp c$. A relatively small value of $\chi_0^{H\parallel c}=1.8\times 10^{-3}$ emu/Oe Yb-mol and $\chi_0^{H\bot c}=0.4\times 10^{-3}$ emu/Oe Yb-mol is obtained, which is likely attributed to the contributions of core diamagnetism, the van Vleck paramagnetism and Pauli paramagnetism. The derived effective magnetic moments are close to the value of 4.54$\mu_B$ expected for free Yb$^{3+}$ with $J=7/2$, which indicates localized magnetic moments in YNP at high temperatures and the negative values of $\theta_p$ indicate the presence of antiferromagnetic interactions.

In contrast, the magnetic susceptibility of YNA shows a weaker temperature dependence, and the value of $\chi$ is about one order of magnitude smaller than YNP. It can be seen that both $\chi^{H\bot c}$ and $\chi^{H\parallel c}$ start to decrease upon reducing the temperature before displaying an upturn at low temperature, which is likely to originate from a small impurity of Yb$_2$O$_3$, as detected in the XRD measurements. A broad maximum in $\chi(T)$  at high temperatures is commonly observed in IV compounds, as seen for example in YbCuGa and CeRhSb.\cite{Malik1991,Adroja1990} The inset of Fig.~\ref{fig3}(b) indeed shows that a broad peak in the magnetic susceptibility is observed around 400~K, instead of Curie-Weiss behavior. A similar feature was also inferred from measurements of polycrystalline samples up to 350~K \cite{Cho} and this suggests YNA is an IV compound.

\begin{figure}[t]
     \centering
     \includegraphics[width=0.6\linewidth]{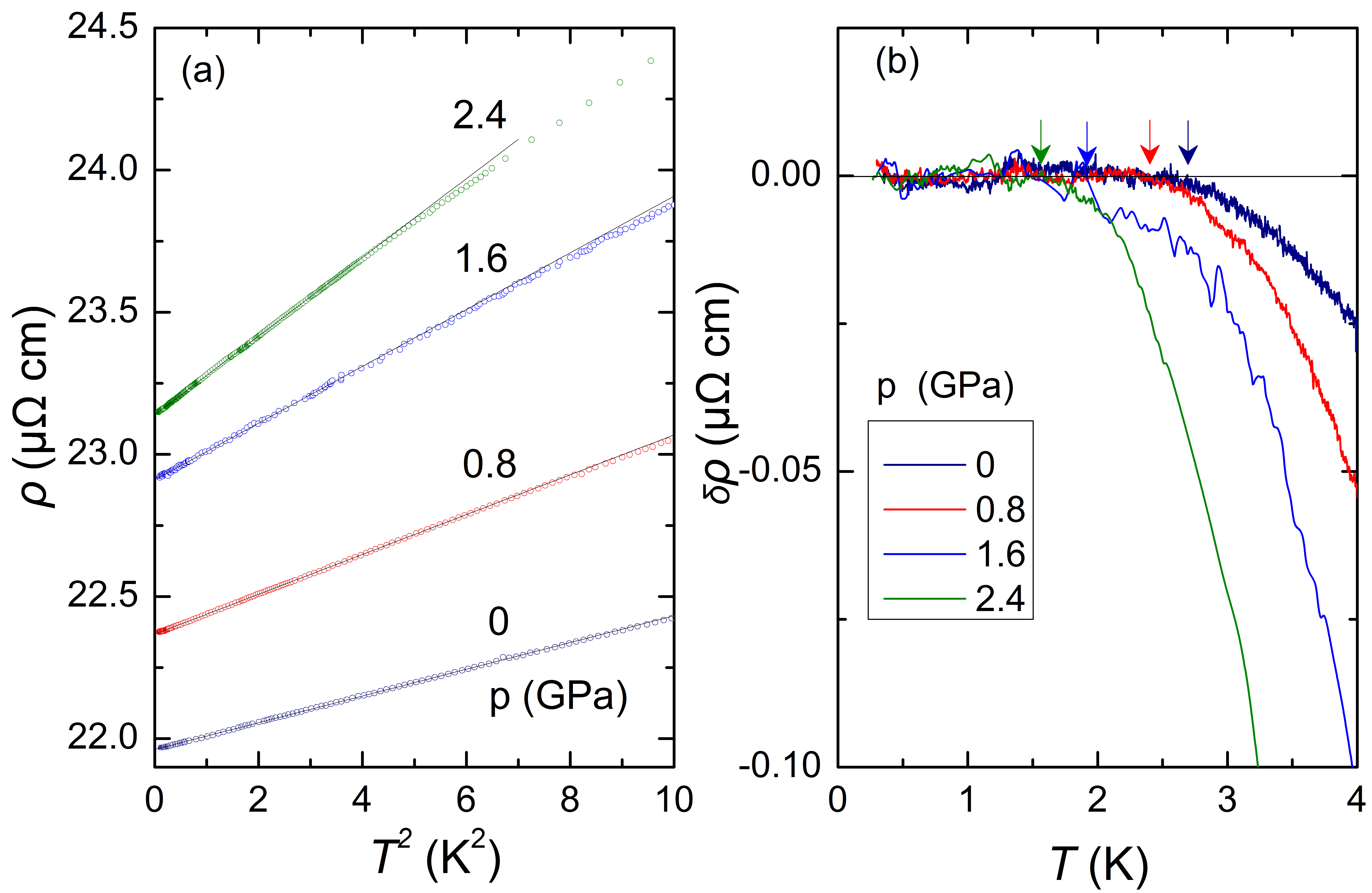}
     \caption{(Color online) (a) Quadratic temperature dependence of the electrical resistivity $\rho(T)$ of single crystal Yb$_2$Ni$_{12}$P$_7$ at several applied pressures up to 2.4~GPa. The solid lines are linear fits to the data. (b) Temperature dependence of  $\Delta\rho(T)$, defined as $\rho(T)-(\rho_0+A T^2)$. The arrows indicate $T_{FL}$, the temperature above which there is a deviation from Fermi liquid behavior.}
     \label{fig4}
\end{figure}

\begin{figure}[t]
     \centering
     \includegraphics[width=0.4\linewidth]{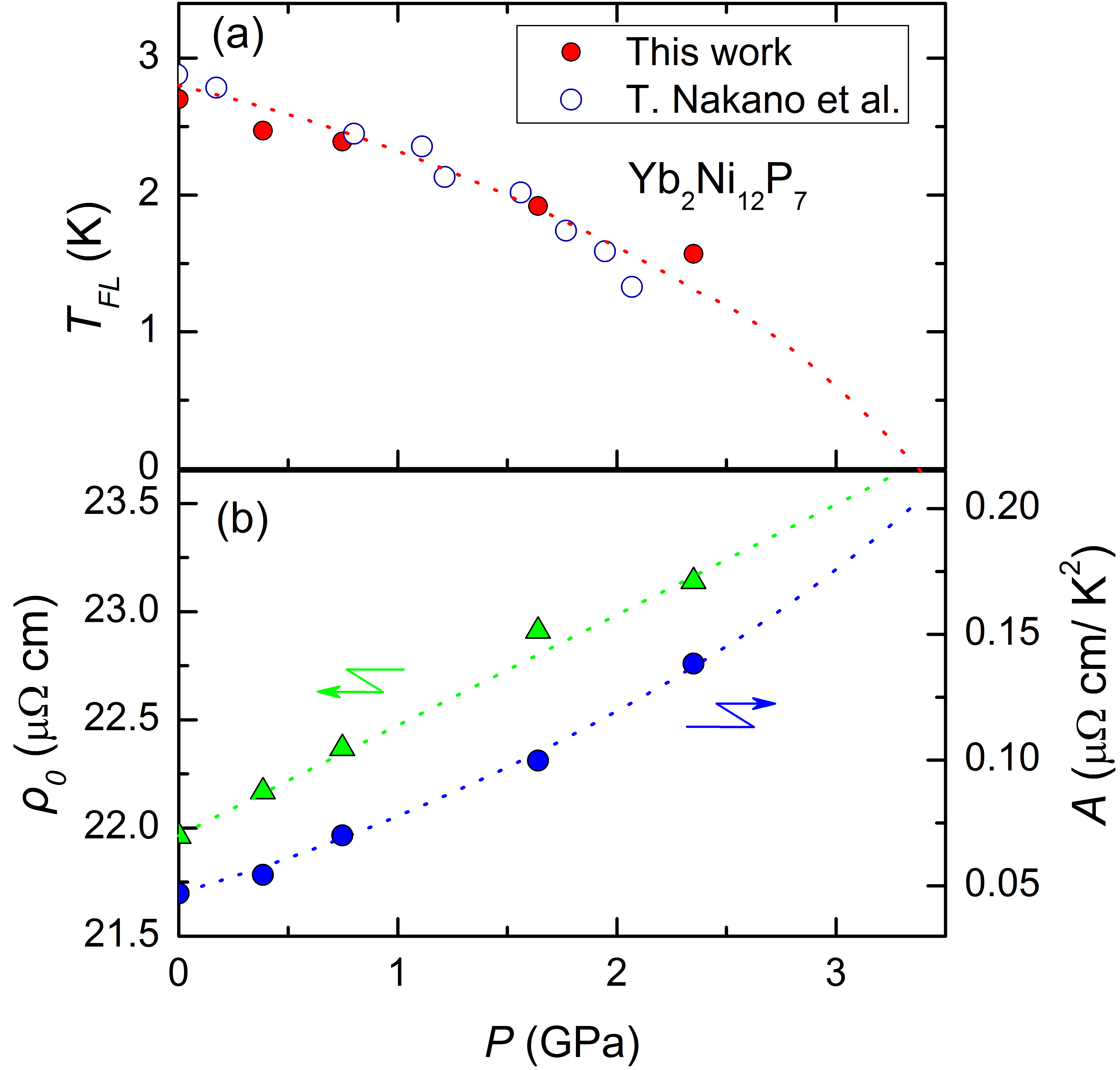}
     \vspace{-12pt}
     \caption{(Color online) Pressure dependence of (a) $T_{FL}$, the temperature below which Fermi liquid behavior is observed in the electrical resistivity of Yb$_2$Ni$_{12}$P$_7$. The results from this work and Ref.~\citeonline{Tomohito} are shown by the solid and open circles respectively. (b) The $A$ coefficient and residual resistivity $\rho_0$, both obtained from fitting the low temperature resistivity with $\rho=\rho_0+A T^2$.}
     \label{fig5}
\end{figure}

\subsection*{Resistivity under pressure}

In order to explore the possibility of a pressure-induced QCP in YNP, we measured the resistivity $\rho(T)$ of a single crystal of YNP at various pressures. Up to a pressure of 2.4~GPa, no phase transition is observed down to about 0.3~K and $\rho(T)$ follows Fermi liquid behavior at low temperatures. As shown in Fig.~\ref{fig4}(a), $\rho(T)$ can be well fitted by $\rho(T)=\rho_0+AT^2$ below the Fermi liquid temperature $T_{FL}$. The values of $T_{FL}$ are determined by fitting the low-temperature resistivity with $\rho(T)=\rho_0+AT^2$ and self-consistently calculating the deviation from Fermi-liquid behavior $\Delta\rho(T)=\rho(T)-(\rho_0+AT^2)$, as shown in Fig.~\ref{fig4}(b). As displayed in Fig.\ref{fig5}(a), the value of $T_{FL}$ decreases with pressure and is extrapolated to zero at a pressure greater than 3~GPa. These results have been plotted alongside those obtained in Ref.\citeonline{Tomohito}, providing strong evidence for the suppression of Fermi liquid behavior with pressure which suggests the existence of a QCP. The increase of the $A$ coefficient and $\rho_0$ with pressure, shown in Fig.\ref{fig5}(b), is evidence for increased Kondo scattering and enhanced electronic correlations, which also indicates the presence of a pressure-induced QCP. We note that the resistive hump around 100K is slightly shifted to lower temperatures with increasing pressure, providing further evidence for an enhancement of Kondo interactions. Magnetic order may occur at higher pressures beyond the QCP, which needs to be confirmed by measurements at higher pressures and lower temperatures.

\begin{figure}[t]
     \begin{center}
     \includegraphics[angle=0,width=0.4\linewidth]{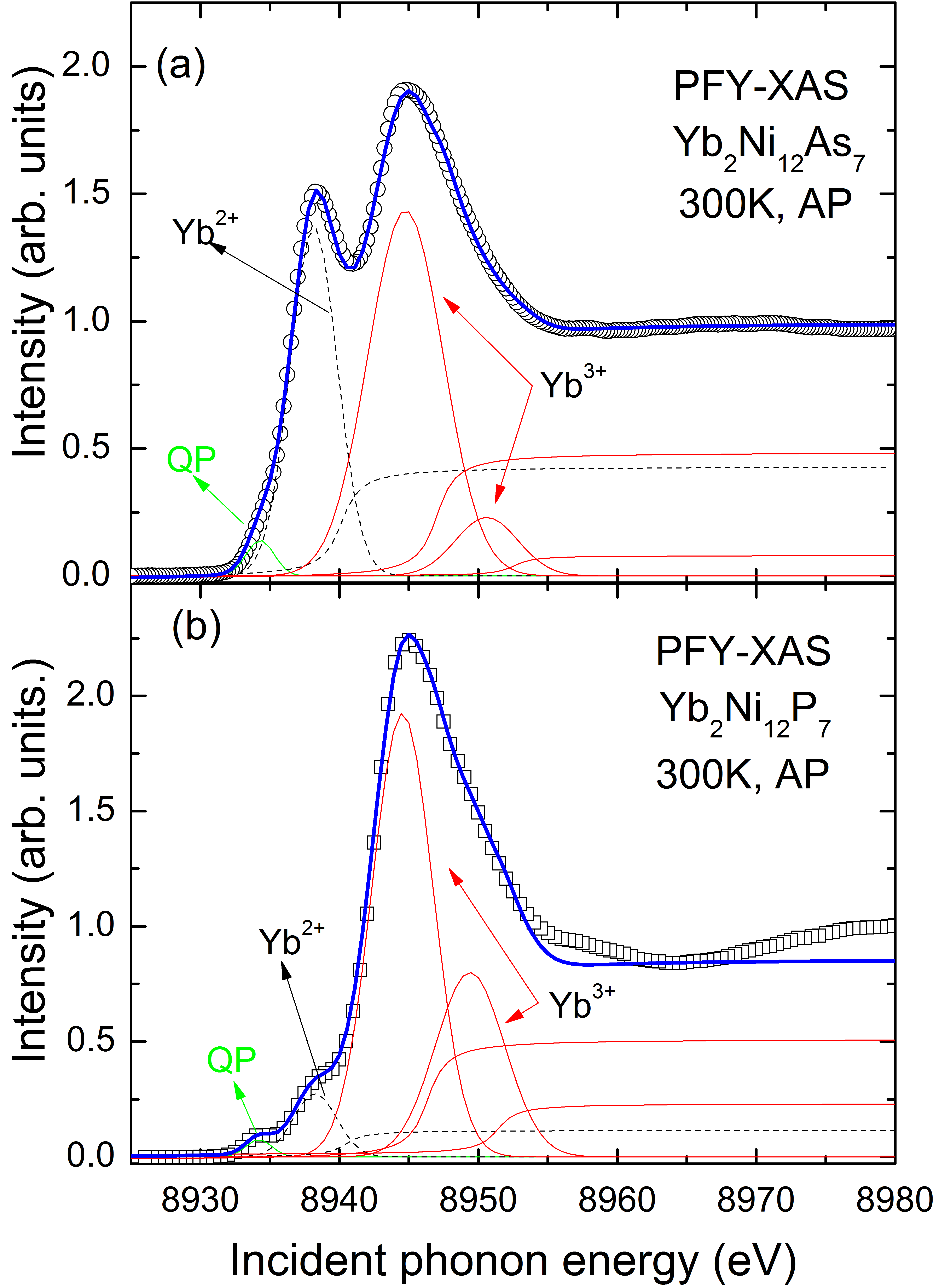}
     \end{center}
     \vspace{-12pt} \caption{(Color online)  Yb-L$_3$ PFY-XAS spectra at room temperature and ambient pressure for polycrystalline (a) Yb$_2$Ni$_{12}$As$_7$ and (b) Yb$_2$Ni$_{12}$P$_7$. The blue solid lines show fits to the data described in the text. The contributions of the quadrupole (QP), Yb$^{2+}$, Yb$^{3+}$ components are shown by the green, dashed and red lines, respectively.}
     \label{fig6}
\end{figure}

\begin{figure*}[t]
     \includegraphics[angle=0,width=\linewidth]{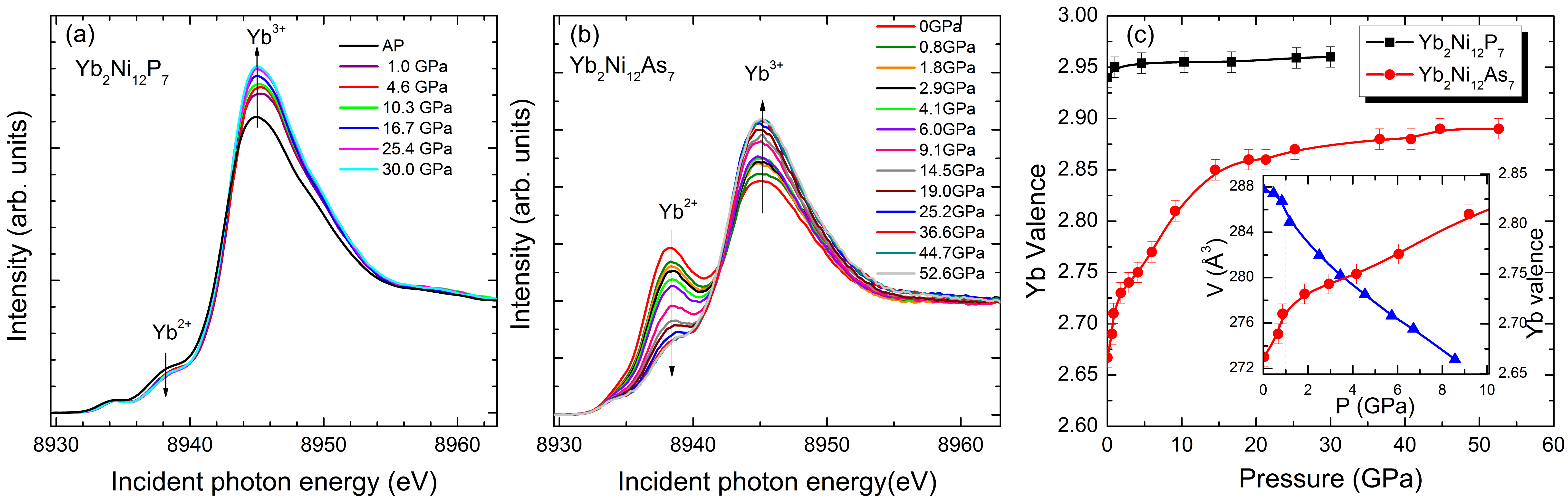}
     \vspace{-12pt} \caption{(Color online) Yb-L$_3$ edge PFY-XAS spectra measured at various pressures for polycrystalline (a) Yb$_2$Ni$_{12}$P$_7$ and (b) Yb$_2$Ni$_{12}$As$_7$. (c) Pressure dependence of the Yb valence for both compounds obtained from fitting the spectra. Here the  systematical error bar of the Yb valence as a function of pressure is estimated to be about $\pm 0.01$. The inset of Fig.~\ref{fig7}(c) shows the pressure dependence of the unit cell volume (blue triangles) and Yb valence (red circles) for YNA. The dashed line at 1~GPa indicates where there is an anomaly in both these quantities. The solid lines are a guide to the eye.}
     \label{fig7}
\end{figure*}

\begin{figure*}[t]
\includegraphics[angle=0,width=\linewidth]{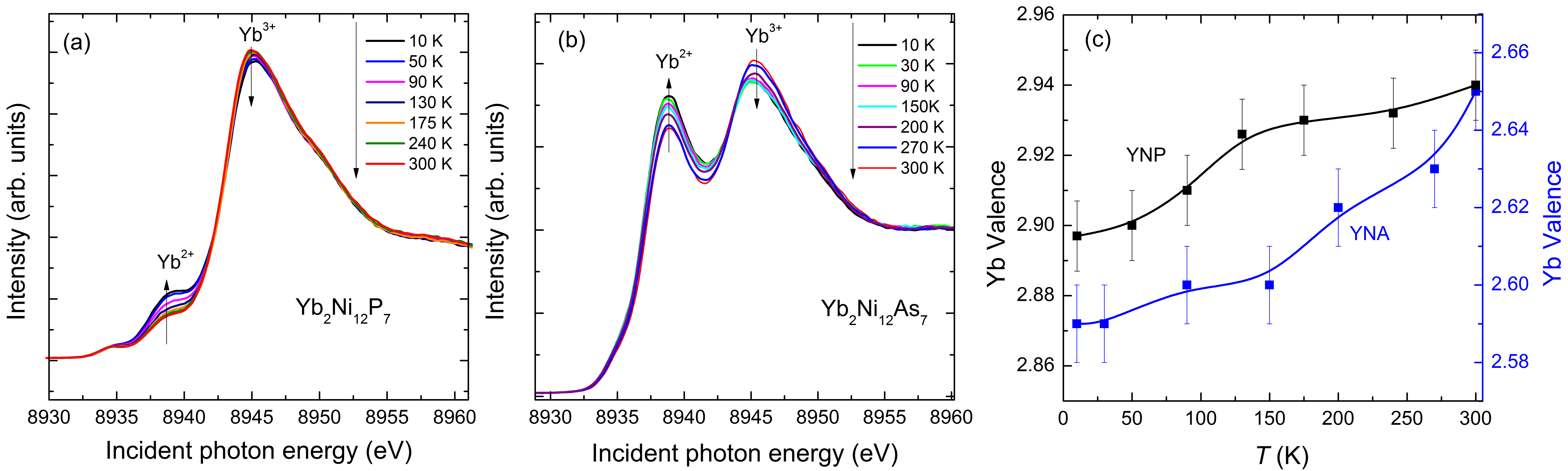}
\vspace{-12pt} \caption{ (Color online)  Yb L$_3$-edge PFY-XAS spectra measured at various temperatures for polycrystalline (a) Yb$_2$Ni$_{12}$P$_7$ and (b) Yb$_2$Ni$_{12}$As$_7$. (c) Temperature dependence of the Yb valence for both compounds obtained from fitting the spectra. The solid lines are a guide to the eye and the same error bar, as determined in Fig. 7, is applied.}
     \label{fig8}
\end{figure*}

\subsection*{X-ray absorption spectroscopy}

Figures~\ref{fig6}(a) and (b) show the Yb L$_3$-edge PFY-XAS spectrum of YNA and YNP in ambient conditions, respectively. Two prominent peaks are observed at ~8938~ev and ~8945~eV, which originate from the $2p_{3/2}\longrightarrow5d$ transitions of Yb$^{2+}$ (4f$^{14}$) and Yb$^{3+}$(4f$^{13}$), respectively. There is a weak shoulder at the lower energy side of the Yb$^{2+}$ peak, at around 8934~eV, which is ascribed to the quadrupole (QP) transition from the Yb $2p_{3/2}$ core to the empty Yb$^{3+}$ $4f$ states due to hybridization between $4f$ and $5d$ electrons. The presence of two significant Yb L$_3$ peaks clearly indicates that there are strong valence fluctuations in YNA. In contrast, the Yb L$_3$ spectrum of YNP mainly consists of the dominant Yb$^{3+}$ component with a weaker Yb$^{2+}$ peak, as shown in Fig.~\ref{fig6}(b), indicating only slightly mixed valence behavior in this compound.

In order to determine the Yb valence of YNP and YNA, the PFY-XAS spectra were fitted with Voigt functions for the QP, Yb$^{2+}$ and Yb$^{3+}$ components along with corresponding arctan-like functions for the edge jump. To fit the data, two components are necessary to account for the absorption features associated with Yb$^{3+}$, which can be attributed to the crystal field splitting of Yb $5d$ band, similar to that observed in other Yb-based intermediate valence compounds such as YbCuAl, \cite{Hitoshi2013} which has a clear double-peak structure. The mean Yb valence ($\nu$) can be estimated to be +2.94 and +2.65 for YNP and YNA respectively using $\nu=2+I(3+)/[I(3+)+I(2+)]$, where $I(2+)$ and $I(3+)$ are the respective intensities of the Yb$^{2+}$ and Yb$^{3+}$ components.

Figures~\ref{fig7}(a) and (b) show the pressure dependence of the Yb L$_3$-edge PFY-XAS spectra of YNP and YNA respectively. The spectral weight is transferred from the Yb$^{2+}$ to Yb$^{3+}$ peaks with increasing pressure, indicating that at sufficiently high pressure, the Yb valence approaches +3. The pressure dependence of the Yb valence is displayed in Fig.~\ref{fig7}(c). For YNP (black squares), the Yb valence increases slightly from +2.94 at ambient pressure to +2.96 at 30~GPa, and therefore gradually approaches +3 at high pressure. In contrast, there is a significant increase of the Yb valence in YNA (red circles) upon applying pressure, from +2.66 when the pressure cell is in ambient conditions to +2.85 at around 15~GPa. We note that the Yb valence of YNA in ambient conditions shows a slight difference between measurements performed with the sample inside and outside of the diamond anvil cell, which is due to different background contributions from the fluorescence signal. For pressures greater than 15~GPa, the slope of the valence increase becomes smaller and the valence increases by about 0.1 up to 50~GPa. In the low pressure region for YNA, which is shown in the inset of Fig.~\ref{fig7}(c), a clear feature is observed at around 1~GPa, where there is a sudden change in the slope of the Yb valence. This is displayed alongside the pressure dependence of the unit cell volume from Fig.\ref{fig1}(b) and it can be seen that the sharp drop in the cell volume around 1 GPa (blue triangles) coincides with the feature in the Yb valence, which may indicate a weak volume collapse transition in YNA.

The temperature dependence of the Yb L$_3$-edge PFY-XAS spectra for YNP and YNA was also measured from room temperature to 10~K, as displayed in Fig.~\ref{fig8}. With decreasing temperature, the intensity of the Yb$^{2+}$ component increases at the expense of the Yb$^{3+}$ component in both compounds, indicating enhanced valence fluctuations at low temperatures. The temperature dependence of the Yb valence was estimated from fitting the spectra and the results are shown in Fig.~\ref{fig8}(c). The Yb valence decreases from +2.94 at room temperature to +2.90 at 10~K for YNP, while the corresponding decrease in YNA is from +2.65 to +2.59. These results indicate a relatively weak temperature dependence of the Yb valence in both compounds. Below 130~K, there is a stronger decrease in the valence of YNP, which is consistent with the behavior of many Yb-based heavy fermion compounds, \cite{Yamaoka2009} where the valence decreases more rapidly upon the onset of Kondo screening. The Yb valence of YNP, even at low temperatures is only slightly below +3, the value where the 4f shell has been depopulated of one electron. A more strongly mixed valence state with an Yb valence of +2.79 at room temperature was previously obtained from analyzing the magnetic susceptibility with an ICF model. \cite{Cho,Jang} However, this is an empirical model where the susceptibility has contributions from the Yb$^{3+}$ and Yb$^{2+}$ ions. As a result of the hybridization between the $4f$ and conductions electrons, there may be a significant screening of the magnetic moment, which will lead to a reduced contribution from the  Yb$^{3+}$ in the ICF model, even if the valence remains nearly +3.

\begin{figure}
  \begin{center}
     \includegraphics[angle=0,width=0.5\linewidth]{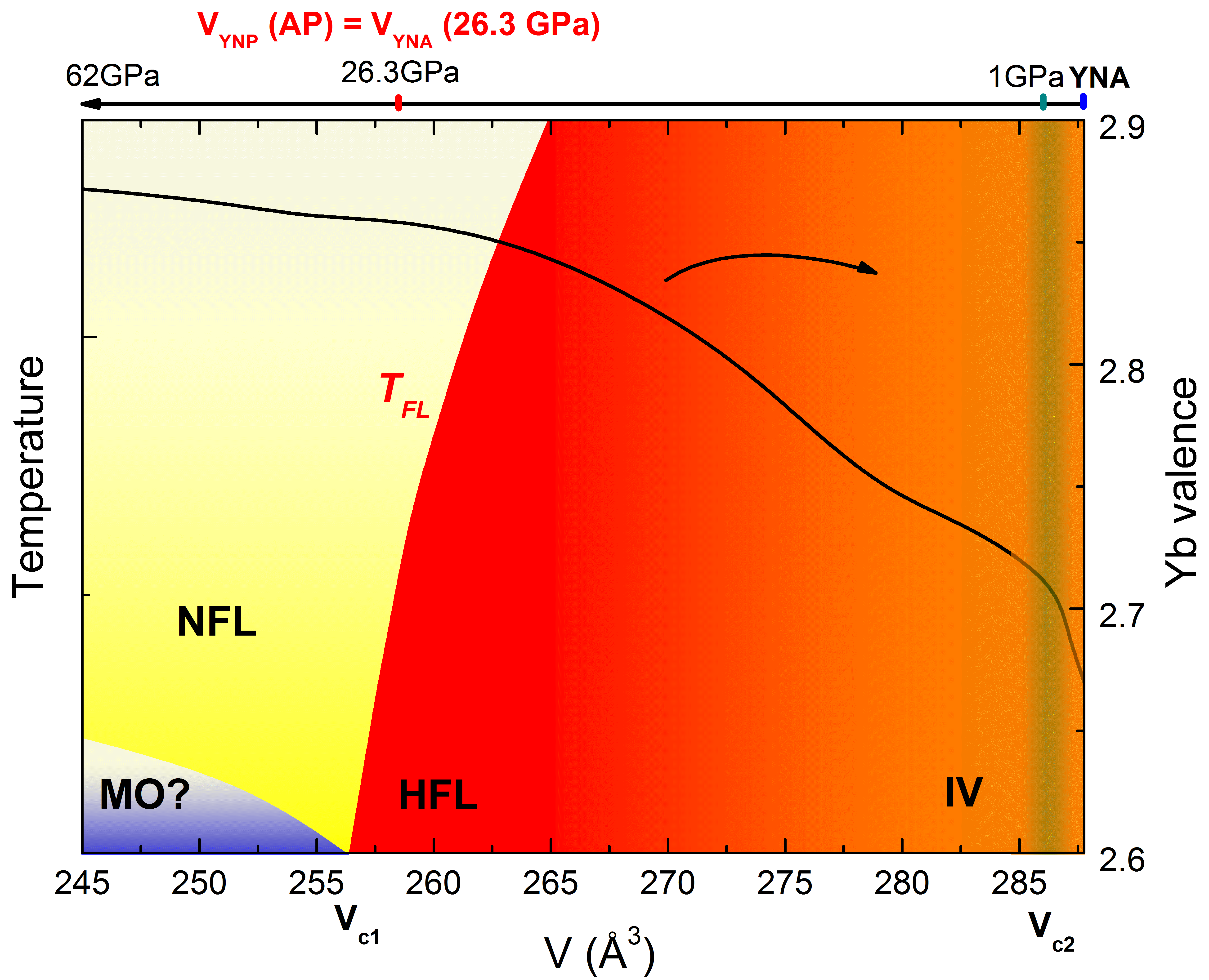}
  \end{center}
     \vspace{-12pt} \caption{(Color online) Schematic phase diagram of Yb$_2$Ni$_{12}$Pn$_7$($Pn$=P, As). Yb$_2$Ni$_{12}$As$_7$ is in the intermediate valence (IV) regime at ambient pressure, but the Yb valence increases and the system becomes a heavy Fermi liquid (HFL) upon reducing the unit cell volume. The line marked $T_{FL}$ shows the temperature where there is an onset of Fermi liquid behavior at low temperatures. A magnetic QCP is expected to occur at $V_{c1}$ (the position in the figure corresponds to a pressure of 3.5~GPa for YNP). Beyond this point a magnetically ordered (MO) phase may occur but this awaits experimental confirmation. The region labelled $V_{c2}$ corresponds to where an anomaly is observed in both the volume and Yb valence of Yb$_2$Ni$_{12}$As$_7$. The solid black line shows the volume dependence of the Yb valence of Yb$_2$Ni$_{12}$As$_7$. }
     \label{fig9}
\end{figure}

\section*{Discussion and summary}

Our results are summarized in the schematic temperature-volume phase diagram of the Yb$_2$Ni$_{12}Pn_7$ system, displayed in Fig.~\ref{fig9}. The pressure dependence of the unit cell volume of YNA was fitted, as shown in Fig.~\ref{fig1}(c). From extrapolating it to higher pressures, the unit cell volume of YNA  at 26.3~GPa is equal to that of YNP at ambient pressure, as marked on the top of Fig.~\ref{fig9}. Our results show that YNA, which has the larger unit cell is an intermediate valence compound, as indicated by a broad peak in the magnetic susceptibility at around 400~K and the moderately enhanced value of $\gamma=37.4$~mJ/Yb-mol K$^2$. The strongly mixed valence behavior of YNA is confirmed by PFY-XAS,  where the Yb valence at room temperature is +2.65. The weak temperature dependence of the YNA valence indicates that $T_K$ is significantly larger than 300~K, which is consistent with the broad maximum in the magnetic susceptibility at around 400~K.\cite{BauerHF} This is similar to other IV compounds such as YbAl$_3$, where a mixed Yb valence is determined from photoemission spectroscopy measurements \cite{YbAl3a,YbAl3b} and the magnetic and thermodynamic properties are consistent with a value of $T_K$ greater than 500~K.\cite{YbAl3c}

Upon applying pressure, the unit cell of YNA is contracted and the Yb valence increases towards the integer value of +3. At higher pressures, there is a considerably weaker dependence and the valence reaches a near constant value of around +2.88 from 25~GPa, up to at least 53~GPa. At low pressure there is a distinct change in the slope of the Yb valence of YNA which coincides with an anomaly in the unit cell volume around $V_{c2}~\sim286~$\AA$^3$. This suggests that changes in the structural parameters are coupled with changes in the valence, being similar to the volume collapse transition between $\gamma$-Ce and $\alpha$-Ce, where there is no change in the crystal structure but the unit cell volume collapses by $\sim14\%$ and this coincides with changes in the occupancy of the 4$f$ electron levels.\cite{CeRIXS,CeRXES} At around 26.3~GPa, the unit cell volume of YNA is extrapolated to be equal to YNP. The Yb valence of YNP is slightly below the fully trivalent value and there is also little pressure dependence, giving further evidence that YNP can be considered to be a high pressure analogue of YNA and there is a clear crossover from an IV state to a heavy fermion state between the two compounds. An increase in the valence with temperature is observed in PFY-XAS measurements, from around +2.90 at 10~K to +2.94 at 300~K with a slope change at around 130~K in YNP. This reduction of the Yb valence with temperature can be understood in terms of the Anderson impurity model, which leads to an increase of the Yb$^{2+}$ component at low temperatures due to the Kondo effect. \cite{Bickers}

Unlike YNA, the physical properties of YNP are consistent with heavy fermion behavior, as shown by the large specific heat coefficient, the value of the Kadowaki-Woods ratio $R_{KW}$ and the near integer value of the Yb valence. The electrical resistivity of YNP shows non-Fermi-liquid (NFL) behavior at higher temperatures, but Fermi liquid behavior of $\rho=\rho_0+AT^2$ is recovered below $T_{FL}$. As shown in the schematic phase diagram, the Fermi-Liquid temperature $T_{FL}$ is suppressed with pressure and a magnetic QCP is expected to exist above 3 GPa, at around $V_{c1}$. On the other hand, we observe little pressure dependence of the Yb valence in this region for either YNP or the low pressure analogue YNA, and the weak mixed valence behavior might persist far beyond crossing $V_{c1}$. This seems to be a different situation to that recently reported in YbNi$_{3}$Ga$_{9}$,\cite{YbNi3Ga9} where upon applying pressure there is a first-order transition to a magnetically ordered state coinciding with a crossover of the Yb valence to nearly +3. There are in fact several examples of Yb based systems where the valence increases crossing the QCP but displays a much weaker dependence at higher pressures, reaching a near constant value slightly below trivalency. \cite{Yamaoka2010,YbCu2Si2,YbNiGe3}

The above results suggest that any magnetic QCP in the Yb$_2$Ni$_{12}Pn_7$ system is well separated from significant changes in the valence and, therefore, it may be a promising system to separately study the effects of spin and valence fluctuations in Yb based compounds. Further characterizations are highly desirable in order to study the emergent behaviors near both $V_{c1}$ and $V_{c2}$. For example, careful measurements under higher pressures are needed to determine the existence of a magnetic QCP in Yb$_2$Ni$_{12}Pn_7$. For YNA, it would be of interest to relate the measured changes in the valence with the low temperature behavior, particularly around $V_{c2}$, where there may be evidence of critical phenomena or other unusual properties.

\section*{Methods}

\textbf{Sample synthesis}  High quality YNA single crystals were synthesized using NiAs self-flux method. Precursors of YbAs, NiAs and Ni were combined in a molar ratio of $2:9:7$ and placed in an alumina crucible, which was sealed in an evacuated quartz tube. The mixture was heated to 1000$^\circ$C, then slowly cooled to 500$^\circ$C at a rate of 3.75$^\circ/$h. Needle-like single crystals with a typical length ~2mm were isolated from the excess flux. YNP single crystals were prepared by using a Sn flux.\cite{Nakano2012} YbP was first prepared by heating Yb ingot and P powder at 1050$^\circ$C for two weeks. Precursors of YbP, Ni, P and Sn were combined in an alumina crucible in the molar ratio of $2:12:5:10$  and sealed in an evacuated quartz tube. The mixture was heated to 1000$^\circ$C where it was held for 10 hours before being cooled to 800$^\circ$C at 4$^\circ/$h. The crystals were separated from the flux using a centrifuge and any remaining flux was removed by etching in HCl. High quality polycrystalline samples of YNP (YNA) were also synthesized using a solid state reaction method. Stoichiometric quantities of  Yb ingot, Ni powder and P powder (As pieces) were combined and slowly heated to 800$^\circ$C for 240 hours. After being sintered, the sample was ground, pressed into the pellets and annealed at 750$^\circ$C for a week, before being quenched in water.

\textbf{Crystal structure}  The crystal structure of the polycrystalline samples were characterized by room temperature powder x-ray diffraction (XRD) using a PANalytical XPert MRD diffractometer with Cu K$\alpha1$ radiation and a graphite monochromator. The structural refinement was carried out using the GSAS+EXPUI software. \cite{Rietveld,Toby2001} The high pressure powder XRD measurements were performed at BL12B2 Taiwan beamline of SPring-8. A finely grained powder of YNA along with dispersed tiny ruby balls used to determine the pressure, were loaded into the sample chamber (diameter 235$\mu$m) of a stainless steel gasket, which is mounted on a Boehler-Almax Plate diamond anvil cell (DAC) with a culet size of 450$\mu$m. The pressure-transmitting medium was a mixture of methanol, ethanol and water in the ratio of $16:3:1$. \cite{Jacobsen}
The hydrostaticity of the applied pressure was confirmed by measurements of the fluorescence line shift of ruby at multiple points before and after each exposure. With a monochromatic beam ($\lambda=0.6199$~\AA), the XRD patterns were recorded by using an ADSC Quantum 4R CCD x-ray detector and then transformed into 1D patterns using the program FIT2D. The experimental setup was calibrated using a high quality CeO$_2$ standard ($99.99\%$, Aldrich), which was used to determine the sample-to-detector distance.

\textbf{Physical properties measurements} The temperature dependence of the resistivity was measured using a standard four probe technique with a Physical Property Measurement System (Quantum Design PPMS-9T). The low-temperature resistivity (down to 0.3~K) was measured in a $^3$He refrigerator. The specific heat measurements were performed in a PPMS-9T using the two-$\tau$ relaxation method. Magnetic susceptibility measurements were performed  down to 2~K  using superconducting quantum interference device (SQUID) magnetometer, Magnetic Property Measurement System (Quantum Design MPMS-5T). The resistivity under pressure was measured in a piston-cylinder-type pressure cell up to 2.4~GPa, the maximum pressure which could be applied. Daphne 7373 was used as the pressure transmitting medium.

\textbf{Hard X-Ray spectroscopy} Yb L$_3$-edge partial-fluorescence-yield x-ray absorption spectroscopy (PFY-XAS) was  measured at the Taiwan inelastic x-ray scattering beamline BL12XU of SPring-8. The undulator beam was monochromated by a pair of Si(111) crystals and focused to an area of 30 $\times$ 30 $\mu$m$^2$ at the sample position with two Kilpatrick-Baez (KB) focusing mirrors. The Yb L$_{\alpha1}$ x-ray emission was collected at 90$^\circ$ from the incident x-rays and analyzed with a spectrometer (Johann type) equipped with a spherically bent Si(620) crystal and a solid-state detector (XFlash 1001 type 1201) arranged on a horizontal plane in a Rowland-circle geometry (radius 1~m). In order to suppress core-hole lifetime broadening effects and obtain the high resolution XAS spectra, the PFY-XAS spectra were collected by monitoring the emission at the maximum intensity of the Yb L$_{\alpha1}$ fluorescence line at 7416~eV. The overall energy resolution was evaluated to be $\sim$1~eV from the full width at half maximum of the quasi-elastic scattering from the sample, which is centred at 7416~eV, the emitted photon energy. The intensities of all spectra were normalized by the incident beam intensity, which is monitored at a position just before the sample. A Mao-Bell diamond anvil cell with a Be gasket was used for the high-pressure XAS measurements. A Be gasket with 3.5 mm diameter, pre-indented to approximately 50 $\mu$m thick was used. The diameter of the sample chamber in the Be gasket was ~100 $\mu$m and the culet size of diamond anvil cell was 350 $\mu$m. Silicone oil was used as the pressure transmitting medium. The applied pressure in the DAC was determined by averaging measurements of the ruby luminescence, before and after measuring each spectrum. A closed-cycle cryostat was used to measure down to 10~K.


\section*{Acknowledgments}
We are grateful to F. Steglich and Han-Oh Lee for their valuable suggestions. This work is partially supported by the National Natural Science Foundation of China (Nos. 11174245 and 11474251), the National Basic Research Program of China (No. 2011CBA00103) and the Fundamental Research Funds for the Central Universities in China.

\section*{Author contributions}

H.Q.Y. planned the experiments. W.B.J, L.Y., Y.F.W, T.S., Z.W.C. and F.G. synthesized the samples. Sample characterization at ambient pressure was performed by W.B.J, L.Y. and Y.F.W., while C.Y.G. and X.L. carried out resistivity measurements under pressure. The synchrotron measurements were performed by Z.H., J.M.L., H.I., K.D.T., Y.F.L., L.H.T. and J.M.C.. The data were analysed by W.B.J., C.Y.G., Z.H., J.M.L., M.S., J.M.C. and H.Q.Y., and W.B.J., Z.H., M.S., J.M.C., and H.Q.Y. wrote the paper. All authors participated in discussions and approved the submitted manuscript.

\section*{Additional information}
\textbf{Competing financial interests:} The authors declare no competing financial interests.

\end{document}